\documentclass[twocolumn,aps,superscriptaddress,showpacs]{revtex4}

\usepackage{amssymb}
\usepackage{amsmath}
\usepackage{graphicx}
\usepackage[normalem]{ulem}
\usepackage[dvips]{color}

\usepackage[normalem]{ulem}  
\usepackage[dvips]{color} 

\renewcommand\sout{\bgroup \color{red} \ULdepth=-.5ex \ULset}

\begin{document}

\title{Mean-field effects on particle and antiparticle elliptic flows in the beam-energy scan program at RHIC}
\author{Jun Xu}
\email{xujun@sinap.ac.cn}
\affiliation{Shanghai Institute of Applied
Physics, Chinese Academy of Sciences, Shanghai 201800, China}
\author{Che Ming Ko}
\affiliation{Cyclotron Institute and
Department of Physics and Astronomy, Texas A$\&$M University,
College Station, Texas 77843, USA}
\author{Feng Li}
\affiliation{Cyclotron Institute and
Department of Physics and Astronomy, Texas A$\&$M University,
College Station, Texas 77843, USA}
\author{Taesoo Song}
\affiliation{Frankfurt Institut for Advanced Studies and Institute for Theoretical Physics,
Johann Wolfgang Goethe Universitat, Frankfurt am Main, Germany}
\author{He Liu}
\affiliation{Shanghai Institute of Applied Physics, Chinese Academy
of Sciences, Shanghai 201800, China}

\date{\today}
\begin{abstract}
The elliptic flow splitting between particles and their
antiparticles has recently been observed by the STAR Collaboration
in the beam-energy scan program at the Relativistic Heavy Ion
Collider. In studies based on transport models, we have found that
this splitting can be explained by the different mean-field
potentials acting on particles and their antiparticles in the
produced baryon-rich matter. In particular, we have shown that the
experimentally measured relative elliptic flow difference can help
constrain the vector coupling constant in the Nambu-Jona-Lasinio
model used in describing the partonic stage of heavy-ion collisions.
This information is useful for locating the critical point in the
QCD phase diagram and thus understanding the phase structure of QCD.
\end{abstract}


\maketitle

\section{Introduction}
\label{introduction}

The main goal of experiments on relativistic heavy-ion collisions is
to study the hadron-quark phase transition or the QCD phase
structure. At top energies of the Relativistic Heavy Ion Collider
(RHIC) and the Large Hadron Collider (LHC), the produced quark-gluon
plasma (QGP) is essentially baryon free and the phase transition is
thus a smooth crossover according to results from the lattice QCD
calculations~\cite{Ber05,Ako06,Baz12}. On the other hand, studies
based on various theoretical models have predicted that the
hadron-quark phase transition becomes a first-order one at large
baryon chemical potential~\cite{Asa89,Fuk08,Car10,Wei12}. A critical
point is thus expected to exist between the smooth crossover and the
first-order phase transition. To search for its signature,
experiments under the beam-energy scan (BES) program have recently
been carried out at RHIC. Although there is no definitive conclusion
on the location of the QCD critical point, some interesting
phenomena different from those observed in heavy-ion collisions at
much higher energies have been observed~\cite{Kum11,Moh11}, such as
the weakening in the charge separation, the net proton number
fluctuation, the fluctuation of $p/\pi$ and $K/\pi$ ratios, the
monotonic decrease of the freeze-out eccentricity with increasing
beam energy from the Hanbury-Brown Twiss analysis, the larger
nuclear modification factor for high-transverse-momentum particles
than in collisions at higher energies, and the splitting of the
direct flow and elliptic flow between particles and their
antiparticles.

The observed elliptic flow ($v_2$) splitting between particles and
their antiparticles~\cite{STAR13} in the BES program at RHIC, which
obviously indicates the break down of the number of constituent
quark scaling established at higher collision
energies~\cite{STAR07}, has attracted much attention. Various
explanations have been suggested for understanding this phenomenon.
It was shown in Ref.~\cite{Bur11} that the chiral magnetic wave
induced by the strong magnetic field from non-central heavy-ion
collisions could lead to a charge quadrupole moment in the
participant region, which would then result in a larger $v_2$ for
positively charged particles than negatively charged ones,
especially for charged pions due to their similar final-state
interactions. Also, the different $v_2$ between particles and their
antiparticles has been attributed to different $v_2$ of produced and
transported particles~\cite{Dun11}, different rapidity distributions
for quarks and antiquarks~\cite{Gre12}, and the conservation of
baryon charge, strangeness, and isospin~\cite{Ste12}.

On the other hand, we have shown in our recent
studies~\cite{Xu12,Son12,Xu14,Ko13,Ko:2014xua} that the observed
$v_2$ splitting can be explained by the different mean-field
potentials acting on particles and their antiparticles. The effect
of the mean-field potential or the nuclear matter equation of state
on the elliptic flow was already known in heavy ion collisions at
SIS energies of the order 1 AGeV~\cite{Dan02}.  At this energy, the
expansion of the hot participant matter is blocked by the cold
spectator matter, resulting in the emission of more particles out of
the reaction plane and thus a negative $v_2$. A more repulsive
mean-field potential leads to a fast expansion of the participant
matter, thus a stronger blocking effect and an even more negative
$v_2$. This becomes different in relativistic heavy-ion collisions
where the blocking effect is absent as the spectator matter quickly
moves away from the participant region. As a result, particles with
repulsive potentials are more likely to leave the system, while
those with attractive potentials are more likely trapped in the
system. For the case of a positive eccentricity of the participant
region in noncentral collisions, the $v_2$ is then enhanced for
particles with repulsive potentials and suppressed for those with
attractive potentials. In the baryon-rich matter formed in heavy-ion
collisions at $\sqrt{s_{NN}}=7.7\sim39$ GeV in the BES program at
RHIC, particles usually have repulsive or less attractive potentials
compared to the strong attractive potentials for their
antiparticles, and the difference becomes smaller with decreasing
net baryon density at higher collision energies. This qualitatively
explains the observed decreasing $v_2$ splitting with increasing
beam energy.

This paper is organized as follows. In Sec.~\ref{ampt}, we briefly
review the physics contents of a multiphase transport model based on
which our studies were carried out. The hadronic potentials and the
partonic potentials as well as their effects on $v_2$ splitting
between particles and antiparticles are discussed in
Sec.~\ref{hadronic} and Sec.~\ref{partonic}, respectively. Results
with the inclusion of both potentials are shown in Sec.~\ref{both}.
Finally, conclusions and outlook are given in Sec.~\ref{summary}.

\section{The AMPT model}
\label{ampt}

Our study of the mean-field effects on $v_2$ splitting between
particles and their antiparticles were based on a multiphase
transport (AMPT) model~\cite{Lin05}. To properly describe the
relative contributions of the partonic phase and the hadronic phase
on the final elliptic flow of hadrons, we use the string melting
version, which converts hadrons produced from initial collisions
into their valence quarks and antiquarks, and was shown to
successfully reproduce the charged particle multiplicity, the
collective flow, and the dihadron correlations at both RHIC and
LHC~\cite{Xu11}.

The initial conditions in the AMPT model are obtained from the
heavy-ion jet interaction generator (HIJING) model~\cite{Xnw91},
where both soft and hard scatterings are included by using the Monte
Carlo Glauber model with nuclear shadowing effects. In the string
melting version, the interaction in the partonic phase is described
by parton-parton elastic scatterings based on the Zhang's parton
cascade (ZPC) model~\cite{Zha98}. After the freeze-out of partons, a
spatial coalescence model is used to describe the hadronization
process with the hadron species determined by the flavor and
invariant mass of its constituent quarks and/or antiquarks. The
evolution of resulting hadronic phase is described by a relativistic
transport (ART)~\cite{Bal95}, which has been generalized to include
both baryon-antibaryon annihilations to two-meson states and their
inverse processes~\cite{Lin05}. The cross sections for the former
are determined by the branching ratios for producing corresponding
number of pions according to the phase space considerations, while
the cross sections for the latter are based on the detailed balance
relations. For heavy ion collisions at top energies at RHIC and LHC,
the mean-field potentials for particles and their antiparticles are
not included in either ZPC or ART as they are less important than
partonic and hadronic scatterings on the collision
dynamics~\cite{Xu11}. In
Refs.~\cite{Xu12,Son12,Ko13,Xu14,Ko:2014xua}, the AMPT model was
extended to include mean-field potentials in the hadronic phase and
the partonic phase in order to study heavy ion collisions at
energies carried out in the BES program.

\section{Effects of hadronic potentials}
\label{hadronic}

For nucleon and antinucleon potentials, they can be calculated from
the relativistic mean-field model used in Ref.~\cite{GQL94} to
describe the properties of nuclear matter, i.e.,
\begin{eqnarray}
U_{N,{\bar N}}(\rho_B,\rho_{\bar B})&=& \Sigma_s(\rho_B,\rho_{\bar
B}) \pm\Sigma_v(\rho_B,\rho_{\bar B}),\label{Up}
\end{eqnarray}
where $\Sigma_s(\rho_B,\rho_{\bar B})$ and
$\Sigma_v(\rho_B,\rho_{\bar B})$ are the nucleon scalar and vector
self-energies in a hadronic matter of baryon density $\rho_B$ and
antibaryon density $\rho_{\bar B}$, with "$+$" for nucleons and
"$-$" for antinucleons, respectively. Nucleons and antinucleons
contribute both positively to $\Sigma_s$ but positively and
negatively to $\Sigma_v$, respectively, as a result of the
$G$-parity invariance. The potentials for strange baryons and
antibaryons are reduced relative to those of nucleons and
antinucleons according to the ratios of their light quark numbers.

The kaon and antikaon potentials in the nuclear medium are also
taken from Ref.~\cite{GQL94} based on the chiral effective
Lagrangian, that is $U_{K,{\bar K}} = \omega_{K,{\bar K}} -
\omega_0$ with
\begin{eqnarray}
\omega_{K,{\bar K}} &=& \sqrt{m_K^2 + p^2 - a_{K,{\bar K}}\rho_s
+(b_K\rho_B^{\rm net})^2}\pm b_K\rho_B^{\rm net}\nonumber\\
\end{eqnarray}
and $\omega_0=\sqrt{m_K^2+p^2}$, where $m_K$ is the kaon mass, and
the values of $a_K$, $a_{\bar K}$, and $b_K$ can be found in
Ref.~\cite{GQL97}. In the above, $\rho_s$ is the scalar density and
is calculated self-consistently in terms of the effective quark and
antiquark masses determined from the same relativistic mean-field
model of Ref.~\cite{GQL94} and $\rho_B^{\rm net}=\rho_B - \rho_{\bar
B}$ is the net baryon density. The "$+$" and "$-$" signs are for
kaons and antikaons, respectively.

The pion potentials are related to their self-energies
$\Pi_s^{\pm0}$ according to $U_{\pi^{\pm0}} =
\Pi_s^{\pm0}/(2m_\pi)$, where $m_\pi$ is the pion mass. Only
contributions from the pion-nucleon $s$-wave interaction to the pion
self-energy were included in our study, and they have been calculated up to the
two-loop order in the chiral perturbation theory~\cite{Kai01}. In
asymmetric nuclear matter of proton density $\rho_p$ and neutron
density $\rho_n$, the resulting $\pi^-$ and $\pi^+$ self-energies
are given, respectively, by
\begin{eqnarray}\label{self}
\Pi_s^-(\rho_p,\rho_n)&=&\rho_n[T^-_{\pi N}-T^+_{\pi
N}]-\rho_p[T^-_{\pi N}+T^+_{\pi N}]
\notag\\
&&+\Pi^-_{\rm rel}(\rho_p,\rho_n)+\Pi^-_{\rm cor}(\rho_p,\rho_n),\notag\\
\Pi_s^+(\rho_p,\rho_n)&=&\Pi_s^-(\rho_n,\rho_p).
\end{eqnarray}
In the above, $T^+$ and $T^-$ are, respectively,
the isospin-even and isospin-odd $\pi N$
$s$-wave scattering $T$-matrices, which are given by the one-loop
contribution in chiral perturbation theory; $\Pi^-_{\rm rel}$ is due
to the relativistic correction; and $\Pi^-_{\rm cor}$ is the
contribution from the two-loop order in the chiral perturbation theory.
Their expressions can be found in Ref.~\cite{Kai01}.

For nucleon and strange baryon resonances in a hadronic matter,
we simply extend the above result by treating them as neutron- or proton-like
baryons according to their isospin structure~\cite{Bal95} and light
quark numbers.

\begin{figure}[h]
\centerline{\includegraphics[scale=1.0]{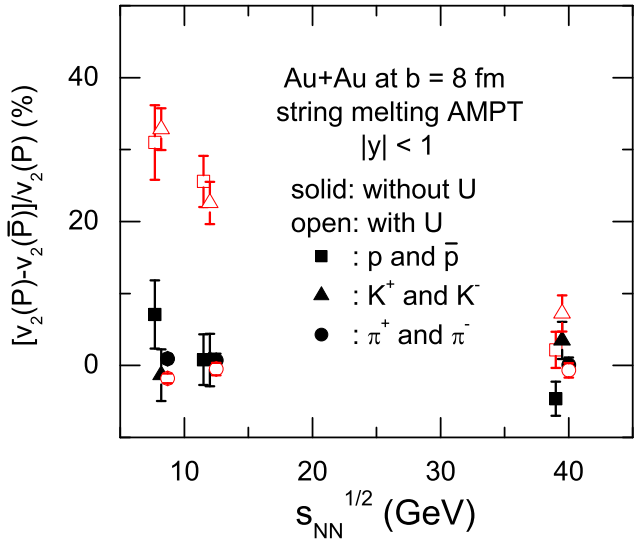}}
\caption{(Color online) Relative elliptic flow difference between
protons and antiprotons, $K^+$ and $K^-$, and $\pi^+$ and $\pi^-$
with and without hadronic mean-field potentials in Au+Au collisions
at $\sqrt{s_{NN}}=7.7$, $11.5$, and $39$ GeV and impact parameter
$\text{b}=8$ fm. Taken from Ref.~\cite{Xu12}.} \label{fig1}
\end{figure}

The mean-field potentials are included in the ART model by using the
test particle method~\cite{Won82}. For the parton scattering cross
section in ZPC and the ending time of the partonic phase, they are
determined from fitting the measured charged particle $v_2$ and
freeze-out energy density calculated from the extracted baryon
chemical potential and temperature at chemical
freeze-out~\cite{And10}. The resulting relative $p_T$-integrated
elliptic flow difference between particles and their antiparticles
at midrapidity ($|y|<1$) are shown in Fig.~\ref{fig1}. It is seen
that without hadronic mean-field potentials, the relative $v_2$
splitting is very small as expected. In the baryon-rich and
neutron-rich matter formed in Au+Au collisions at BES energies, the
hadronic potential is slightly attractive for nucleons, deeply
attractive for antinucleons, slightly repulsive for $K^+$, deeply
attractive for $K^-$, slightly repulsive for $\pi^-$, and slightly
attractive for $\pi^+$. Figure~\ref{fig1} shows that the sign of the
relative $v_2$ splitting is consistent with that expected from the
different hadronic mean-field potentials for particles and their
antiparticles. Also, the $v_2$ difference decreases with increasing
beam energy. These results are qualitatively consistent with the
experimentally measured values of about 63\% and 13\% at $7.7$ GeV,
44\% and 3\% at $11.5$ GeV, and 12\% and 1\% at $39$ GeV for the
relative $v_2$ difference between $p$ and ${\bar p}$ and between
$K^+$ and $K^-$, respectively~\cite{Moh11}. Similar to the
experimental data, the relative $v_2$ difference between $\pi^+$ and
$\pi^-$ is negative at all energies after including their
potentials, although ours have a smaller magnitude.

\section{Effects of partonic potentials}
\label{partonic}

To include mean-field potentials for $u$, $d$, and $s$ quarks and their antiquarks in the partonic phase, we
have developed a partonic  transport model based on the NJL model~\cite{Son12}.
The Lagrangian of the 3-flavor NJL model is given by~\cite{Wei12}
\begin{eqnarray}
\mathcal{L}&=&\bar{\psi}(i\not{\partial}-M)\psi+\frac{G}{2}\sum_{a=0}^{8}\bigg[(\bar{\psi}\lambda^a\psi)^2+(\bar{\psi}i\gamma_5\lambda^a\psi)^2\bigg]\nonumber\\
&+&\sum_{a=0}^{8}\bigg[\frac{G_V}{2}(\bar{\psi}\gamma_\mu\lambda^a\psi)^2+\frac{G_A}{2}(\bar{\psi}\gamma_\mu\gamma_5\lambda^a\psi)^2\bigg]\nonumber\\
&-&K\bigg[{\rm det}_f\bigg(\bar{\psi}(1+\gamma_5)\psi\bigg)+{\rm
det}_f\bigg(\bar{\psi}(1-\gamma_5)\psi\bigg)\bigg],
\end{eqnarray}
with the quark field $\psi=(\psi_u, \psi_d, \psi_s)^T$, the current
quark mass matrix $M={\rm diag}(m_u, m_d, m_s)$, and the Gell-Mann
matrices $\lambda^{a}$ in $SU(3)$ flavor space. In the case that the
vector and axial-vector interactions are generated by the Fierz
transformation of the scalar and pseudo-scalar interactions, their
coupling strengths are given by $G_V=G_A=G/2$, while $G_V=1.1G$ was
used in Ref.~\cite{Wei92} to give a better description of the vector
meson mass spectrum calculated from the NJL model. Other parameters
like $m_u$, $m_d$, $m_s$, $G$, and $K$ are taken from
Refs.~\cite{Wei92,Wei12}. In the mean-field approximation, the above
Lagrangian leads to an attractive scalar mean-field potential for
both quarks and antiquarks. With a nonvanishing $G_V$, it further
gives rise to a repulsive vector mean-field potential for quarks but
an attractive one for antiquarks in a baryon-rich quark matter.

\begin{figure}[h]
\centerline{\includegraphics[scale=0.35]{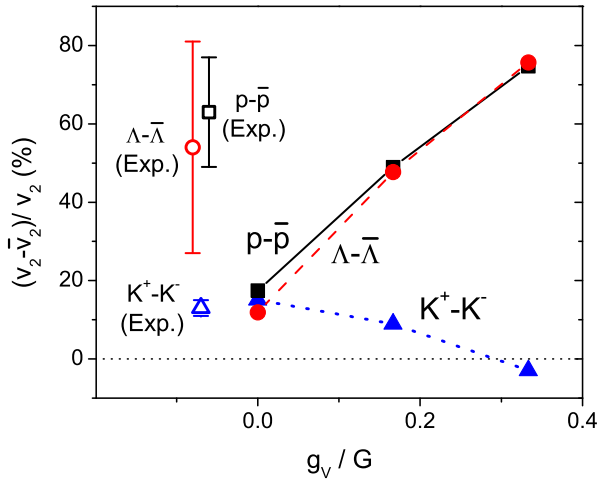}}
\caption{(Color online) Relative elliptic flow difference between
protons and antiprotons, $K^+$ and $K^-$, and $\Lambda$ and
$\bar\Lambda$ as a function of the ratio of the partonic vector
coupling constant to the scalar one in Au+Au collisions at
$\sqrt{s_{NN}}=7.7$ GeV with impact parameter $\text{b}=8$ fm. The
experimental data are taken from Ref.~\cite{Moh11}. Taken from
Ref.~\cite{Ko:2014xua}.} \label{fig2}
\end{figure}

For the initial parton distributions, they are taken from the HIJING
subroutine in the AMPT model. The parton scattering cross section is
determined by fitting the measured $v_2$ of final charged particles.
Again the test particle method is used in the NJL transport model,
and discrete lattices for space are used to calculate the local
density and potential. The evolution of the partonic phase ends when
the energy density of the central cell decreases to about $0.8$
GeV/fm$^3$ that is expected for the quark to hadron phase
transition. The quark matter is then converted to hadrons by the
coalescence model of Refs.~\cite{Greco:2003xt,Chen:2003qj} with the
probability for a quark and an antiquark to form a meson given by
the quark Wigner function of the meson, while the probability for
three quarks or antiquarks to coalesce to a baryon or an antibaryon
given by the quark Wigner function of the baryon or antibaryon. The
relative $v_2$ difference between resulting $p$ and $\bar p$,
$\Lambda$ and $\bar\Lambda$, and $K^+$ and $K^-$ is given in
Fig.~\ref{fig2} as a function of the ratio of the partonic vector
coupling constant to the scalar one in the NJL model, and the
experimental results from Ref.~\cite{Moh11} are also shown for
comparison. Without vector potential, i.e., $g_v=\frac{2}{3}G_V=0$,
there is already slightly $v_2$ splitting between particles and
their antiparticles, and this is due to the slightly larger quark
than antiquark $v_2$ as a result of a smaller initial spatial
eccentricity for quarks than for produced antiquarks. With
increasing $g_v$, the difference between proton and antiproton $v_2$
also increases because of the increasing difference of light quark
and antiquark $v_2$ due to a less attractive potential for quarks
than for antiquarks. Similarly, the relative $v_2$ difference
between $\Lambda$ and $\bar\Lambda$ increases with increasing $g_v$.
On the other hand, the relative $v_2$ difference between $K^+$ and
$K^-$ becomes smaller or even negative for large values of $g_v$.
This is because the vector mean field, which acts similarly on light
and strange (anti-)quarks, leads to a smaller antistrange than
strange quark $v_2$ and also reduces the effect due to different
spatial eccentricities of quarks and antiquarks. For pions, since
the potentials for $u$ and $d$ quarks are the same in the present
3-flavor NJL model due to the neglect of isovector coupling between
quarks and antiquarks, there is no $v_2$ difference between $\pi^+$
and $\pi^-$.

\section{Effects of both hadronic and partonic potentials}
\label{both}

With only hadronic potentials, Figure~\ref{fig1} shows that our
results slightly underestimate the $v_2$ splitting between $p$ and
${\bar p}$ and overestimate that between $K^+$ and $K^-$ in the
experimental data. Comparing the results in Fig.~\ref{fig2} obtained
with only partonic potentials with the experimental data, the
relative $v_2$ difference between protons and antiprotons is still
underestimated, while that between $K^+$ and $K^-$ has a wrong sign.
If the effects from the mean-field potentials on $v_2$ in the
partonic phase and the hadronic phase are additive, one would expect
that the relative $v_2$ splitting between protons and antiprotons as
well as that between $K^+$ and $K^-$ can be quantitatively explained
with both partonic and hadronic potentials for a certain value of
the vector coupling constant in the NJL model. This is indeed the
case as shown in Ref.~\cite{Xu14} by replacing the ZPC model in the
AMPT model with the NJL transport model and ending the partonic
evolution when the effective mass of light quarks in the central
cell increases to about 200 MeV when the chiral symmetry is largely
broken, as well as including the hadronic mean-field potentials in
the ART model.

\begin{figure}[h]
\centerline{\includegraphics[scale=0.85]{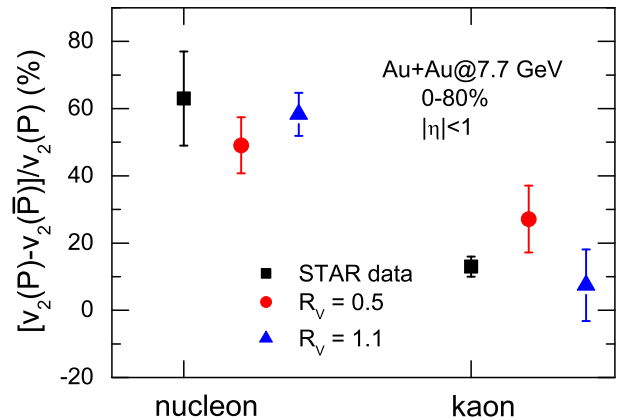}} \caption{(Color
online) Relative elliptic flow difference between protons and antiprotons as well as between $K^+$ and $K^-$ using
two different values for the ratio $R_ v$ of the partonic vector to scalar coupling
constants in mini-bias Au+Au collisions at $\sqrt{s_{NN}}=7.7$ GeV. The
experimental data are taken from Ref.~\cite{Moh11}. Taken from Ref.~\cite{Xu14}.} \label{fig3}
\end{figure}

The relative $v_2$ difference between protons and antiprotons as
well as that between $K^+$ and $K^-$ at midpseudorapidity
($|\eta|<1$) in mini-bias Au+Au collisions at $\sqrt{s_{NN}}=7.7$ in
this more complete study are shown in Fig.~\ref{fig3} for two values
of $R_V=G_V/G=0.5$ and $1.1$. The experimental data from the STAR
Collaboration~\cite{Moh11}, which are shown by filled squares, can
now be quantitatively reproduced with both $R_V=0.5$ and $R_V=1.1$
within the statistical error. Also, we have found that with
increasing value of $R_V$ the relative $v_2$ difference between
protons and antiprotons increases, while that between $K^+$ and
$K^-$ decreases. It is thus expected that further reducing the value
of $R_V$ would underestimate the relative $v_2$ difference between
protons and antiprotons and overestimate that between $K^+$ and
$K^-$, while further increasing the value of $R_V$ would
underestimate that between $K^+$ and $K^-$. To reproduce both the
relative $v_2$ differences between protons and antiprotons as well
as that between $K^+$ and $K^-$ requires the value of $R_V$ to be
about $0.8 \pm 0.3$.

\begin{figure}[h]
\includegraphics[scale=0.4]{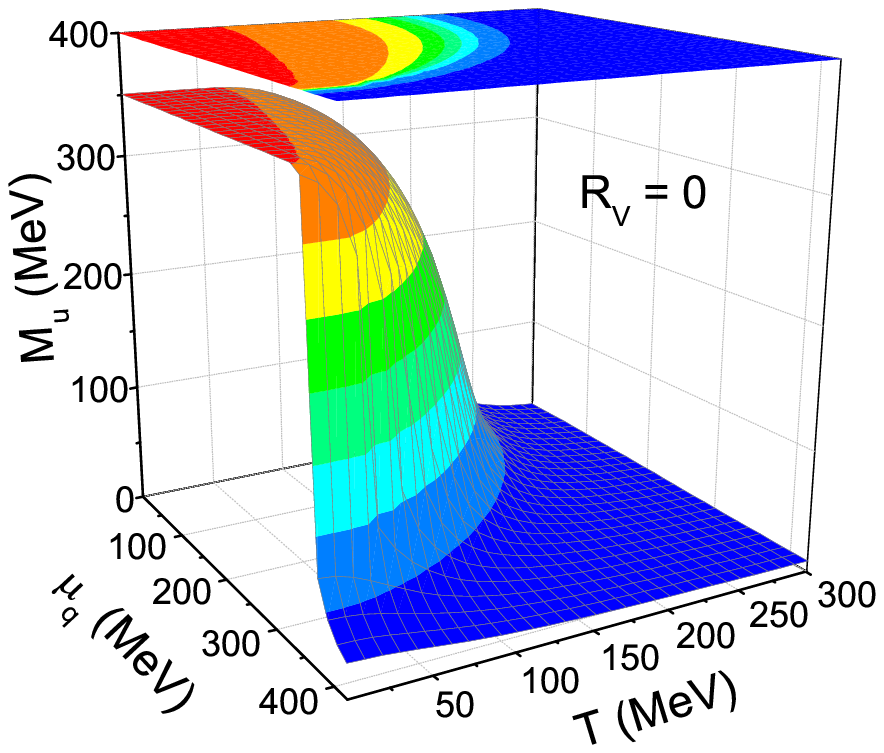}\includegraphics[scale=0.4]{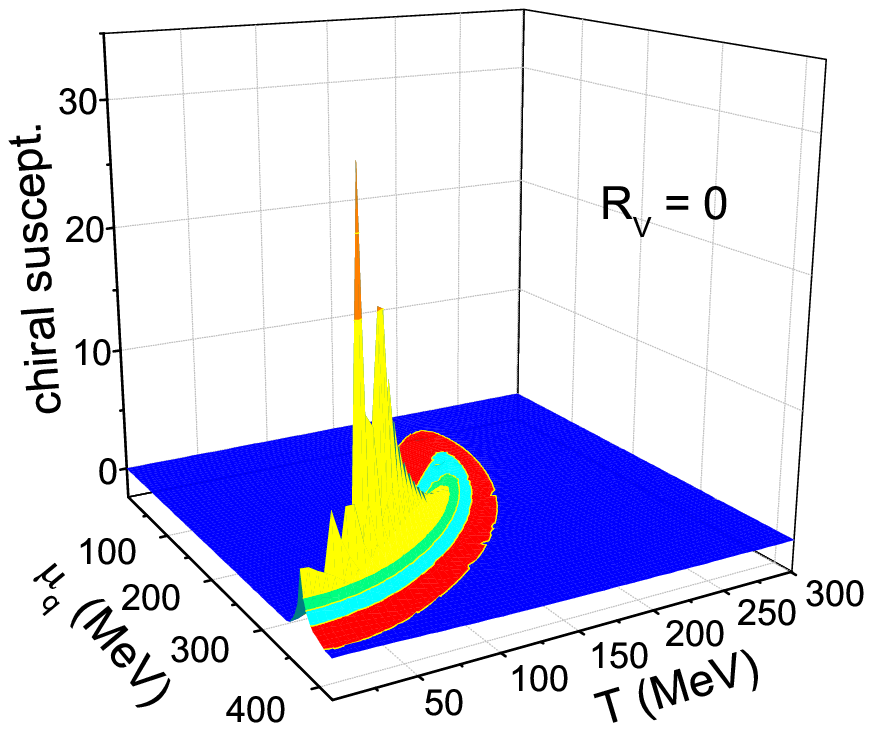}\\
\includegraphics[scale=0.4]{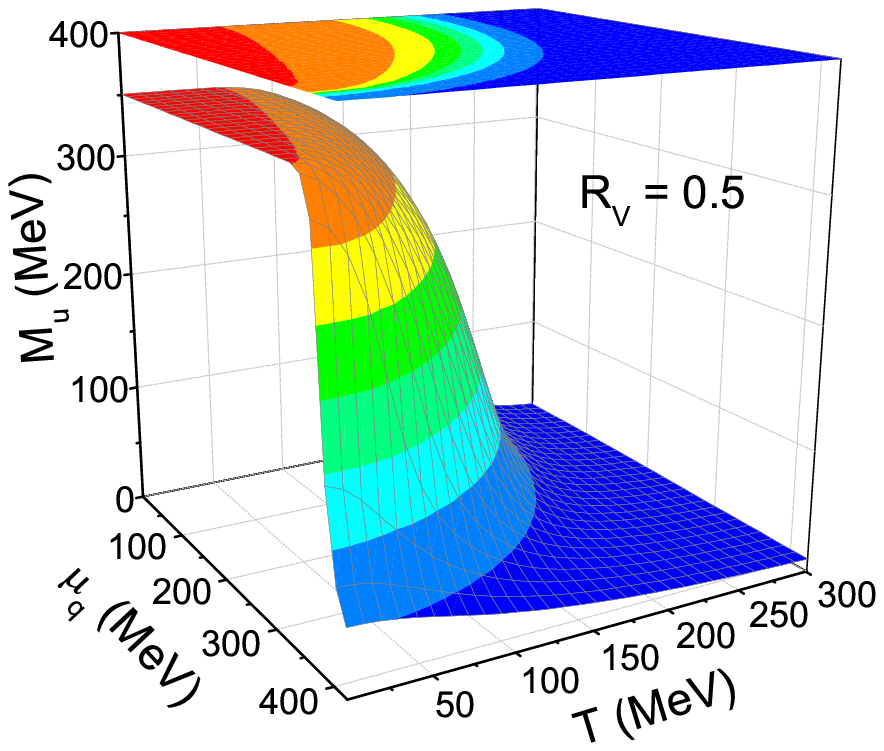}\includegraphics[scale=0.4]{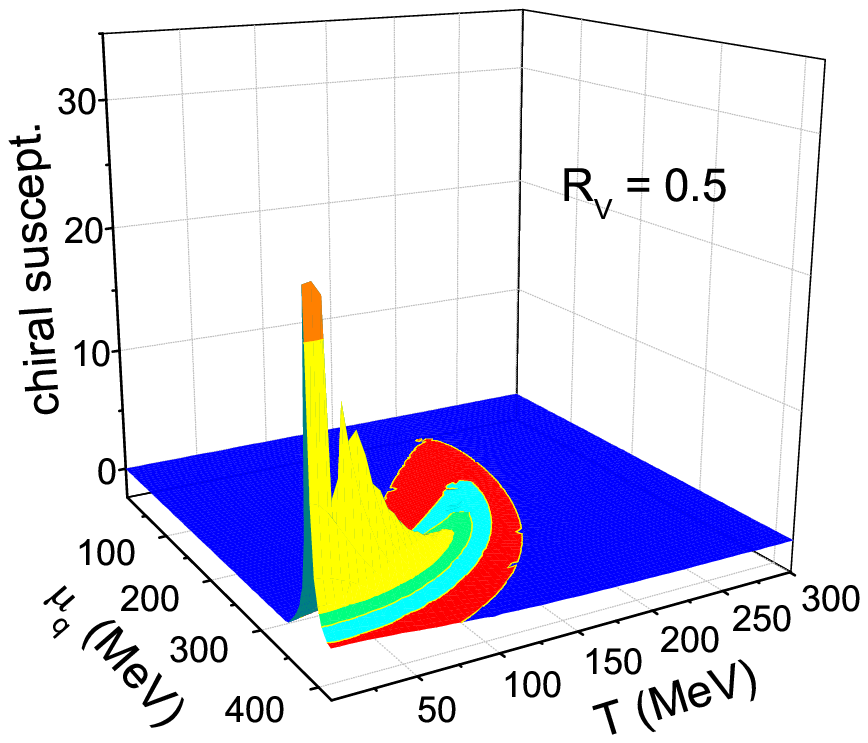}\\
\includegraphics[scale=0.4]{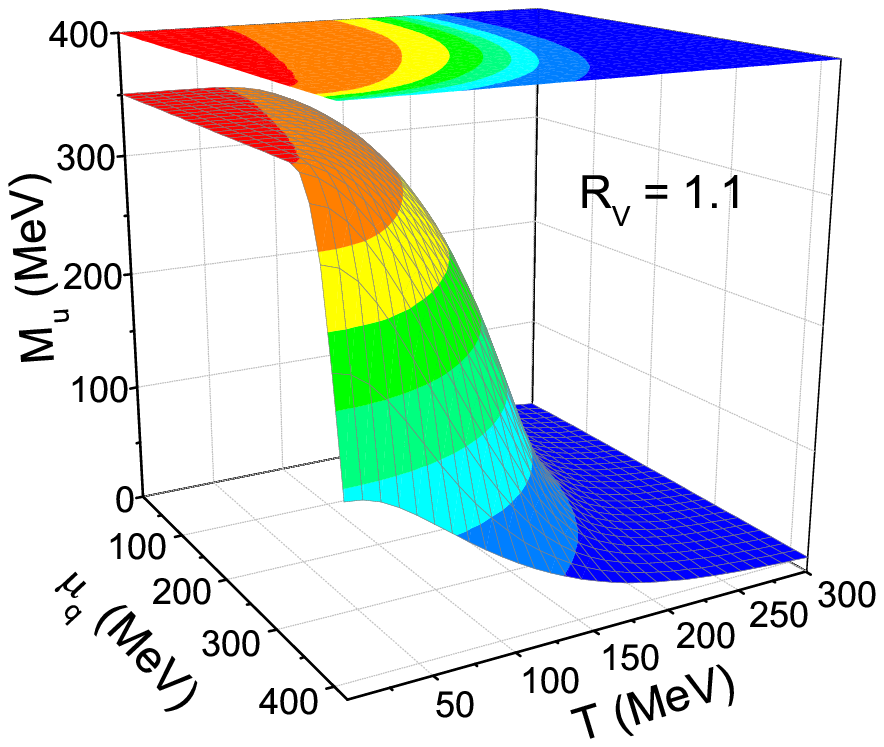}\includegraphics[scale=0.4]{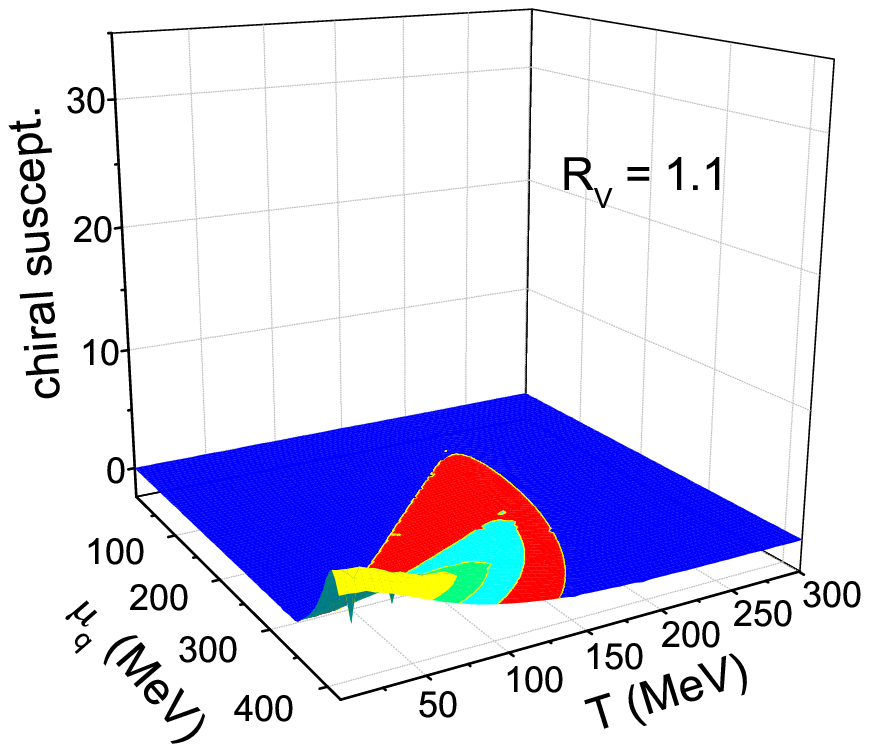}
\caption{(Color online) The dynamical light quark mass (left) and
the corresponding susceptibility (right) in the ($\mu_q$, $T$) plane
from different relative vector coupling constant $R_V=G_V/G$.}
\label{fig4}
\end{figure}

According to Refs.~\cite{Asa89,Fuk08,Car10,Wei12}, the value of the
vector coupling would affect the location of the critical point in
the QCD phase diagram and thus the QCD phase structure. In
Fig.~\ref{fig4} we display the dynamical light quark mass, which is
a function of the quark condensate~\cite{Son12}, and the
corresponding susceptibility, which is defined as the derivative of
the dynamical quark mass with respect to the quark chemical
potential. One sees that at smaller values of quark chemical
potential $\mu_q$ and temperature $T$, the dynamical quark mass is
large, corresponding to the phase where chiral symmetry is broken.
For larger values of $\mu_q$ or $T$, the dynamical quark mass is
similar to the current quark mass, corresponding to the phase where
chiral symmetry is restored. A smooth change from the first phase to
the second phase represents a cross-over chiral phase transition,
while a sudden change represents a first-order phase transition. The
critical point is exactly where the cross-over transition changes to
a first-order one. One sees from the susceptibility that the
critical point moves to lower temperatures with increasing vector
coupling constant, similar to that observed in
Refs.~\cite{Asa89,Fuk08,Car10,Wei12}. This is not surprising because
the equation of state of baryon-rich matter becomes stiff with
larger values of $G_V$, so that the phase transition is more likely
to be a crossover rather than a first-order one.

\begin{figure}[h]
\centerline{\includegraphics[scale=0.85]{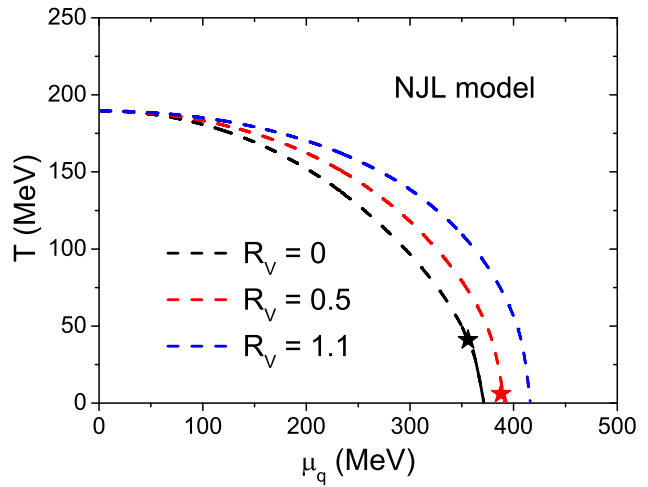}}
\caption{(Color online) The chiral phase boundary and the
corresponding critical point (labeled as stars) based on the
Nambu-Jona-Lasinio model in the ($\mu_q$, $T$) plane from different
relative vector coupling constant $R_V=G_V/G$.}\label{fig5}
\end{figure}

Defining the phase boundary where the chiral condensate is half of
the value at ($\mu_q=0$, $T=0$) as the phase boundary~\cite{Fuk08},
the phase diagram with chiral phase transition with different $R_V$
is shown in Fig.~\ref{fig5}. One should keep in mind that the phase
boundary is only accurate for the first-order phase transition but
is artificially defined for a crossover, as shown by the dashed
lines in Fig.~5. It is thus clearly seen that based on the same NJL
Lagrangian from the transport model study and the phase diagram
calculation, the constraint of the vector coupling $R_V=0.5\sim1.1$
leads to the conclusion that the critical point may disappear or is
located at extremely low temperatures.

\begin{figure}[h]
\centerline{\includegraphics[scale=0.85, bb = 0 0 306
237]{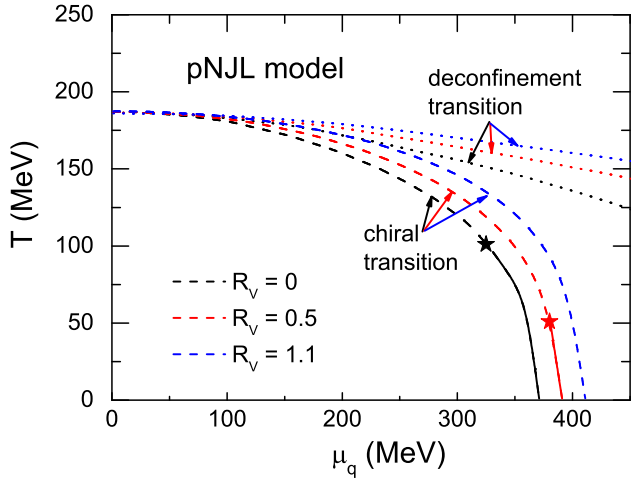}} \caption{(Color online) The chiral phase
boundary, the corresponding critical point (labeled as stars), and
the deconfinement phase boundary based on the Polyakov-looped
Nambu-Jona-Lasinio model in the ($\mu_q$, $T$) plane from different
relative vector coupling constant $R_V=G_V/G$.}\label{fig6}
\end{figure}

The above NJL Lagrangian only includes the contribution of quarks
and antiquarks and can not describe the deconfinement phase
transition. In Ref.~\cite{Fuk04} K. Fukushima included the polyakov
loop representing the gluon contribution, and the poyakov loop can
be taken as the order parameter of the deconfinement phase
transition~\cite{Fuk11}. The polyakov-looped Nambu-Jona-Lasinio
(pNJL) model is identical to the NJL model at zero temperature but
has an additional thermal potential at finite
temperature~\cite{Fuk08}
\begin{equation}
\Omega_{\text{Polyakov}} = -b\cdot T\{54 e^{-a/T} l \bar{l} +
\ln[1-6l \bar{l} - 3(l\bar{l})^2 + 4(l^3+\bar{l}^3)]\},
\end{equation}
where the parameters $a$ and $b$ are fitted to make the quark
condensate and the polyakov loop $l$ cross each other at about
$T=200$ MeV at zero quark chemical potential~\cite{Fuk08}. The phase
structure with the contribution of the polyakov loop is displayed in
Fig.~\ref{fig6}, where the deconfinement phase boundary is defined
as $l=1/2$~\cite{Fuk08}. The deconfinement transition is always a
crossover as a result of finite dynamical quark mass in the pNJL
model. On the other hand, the critical point moves to higher
temperatures after including the polyakov loop, as already discussed
in Ref.~\cite{Fuk08}. However, the conclusion that the critical
point disappears or is located at lower temperatures with larger
values of $R_V$ doesn't change even with the inclusion of the
polyakov loop. It would be of great interest to included gluon
contribution in both the transport model study and the phase diagram
calculation in the future study.

\section{Conclusions and outlook}
\label{summary}

We have reviewed our recent studies on the effects of mean-field
potentials on the elliptic flow splitting between particles and
their antiparticles in the beam-energy scan program at RHIC. By
including the hadronic mean-field potentials in the ART model of a
multiphase transport model, the $v_2$ splitting is qualitatively
consistent with the experimental results, while the magnitude of the
relative $v_2$ difference between protons and antiprotons as well as
$\pi^+$ and $\pi^-$ are underestimated, and that between $K^+$ and
$K^-$ is overestimated. With only partonic potentials from a
3-flavor NJL transport model, the relative $v_2$ difference between
protons and antiprotons is still underestimated while that between
$K^+$ and $K^-$ has a wrong sign when these hadrons are produced
from the coalescence of quarks and antiquarks. Including both the
partonic and hadronic potentials in the AMPT model, the relative
$v_2$ difference between protons and antiprotons as well as that
between $K^+$ and $K^-$ can be quantitatively reproduced, and this
further helps to constrain the $R_V=G_V/G$ in the NJL model to about
$R_V=0.8 \pm 0.3$. Based the NJL Lagrangian used in the transport
model study, we have further calculated the corresponding phase
diagram and found that the resulting critical point in the chiral
phase transition disappears or is located at very low temperatures
based on the obtained constraint of $R_V$, and the conclusion
remains the same even if the contribution of polyakov loop to the
thermal potential is taken into account.

In the previous studies, we have only reproduced the relative $v_2$
difference in mini-bias Au+Au collisions at $\sqrt{s_{NN}}=7.7$ GeV.
It will be interesting to study the $v_2$ splitting at higher
energies as this would provide the possibility of studying the net
baryon or baryon chemical potential dependence of $R_V$. In
addition, since the partonic phase is more important in mid-central
collisions than in peripheral collisions, it will also be of
interest to study the centrality dependence of the $v_2$ splitting
to further constrain the parameters in our model. Since the location
of the QCD critical point is sensitive to the value of $R_V$,
further investigations on the $v_2$ splitting in the baryon-rich
matter can help map out the QCD phase structure.

\begin{acknowledgments}
This work was supported by the National Key Basic Research Program
of China (973 Program) under Contract Nos. 2015CB856904 and
2014CB845401, the "Shanghai Pujiang Program" under Grant No.
13PJ1410600, the "100-talent plan" of Shanghai Institute of Applied
Physics under Grant No. Y290061011 from the Chinese Academy of
Sciences, the National Natural Science Foundation of China under
Grant No. 11475243, the US National Science Foundation under Grant
No. PHY-1068572, and the Welch Foundation under Grant No. A-1358.
\end{acknowledgments}

\end{document}